\documentclass{PoS}
\def\aap{A\&A}                % Astronomy and Astrophysics
\def\aj{AJ}                   % Astronomical Journal
\def\apj{ApJ}                 % Astrophysical Journal
\def\apjl{ApJ}                % Astrophysical Journal, Letters
\def\apjs{ApJS}               % Astrophysical Journal, Supplement
\def\mnras{MNRAS}             % Monthly Notices of the RAS
\def\nat{Nature}              % Nature
\title{A preliminary distance to W\,75N in the Cygnus X star-forming region}

\ShortTitle{Preliminary distance to W\,75N}

\author{\speaker{Kazi L.~J.~Rygl}$^{a,b}$\thanks{Current affiliation IFSI-INAF.}, Andreas Brunthaler$^{a}$, Karl M.~Menten$^a$, Mark J.~Reid$^c$, Huib-Jan van Langevelde$^{d,e}$,Mareki Honma$^f$, Karl J.~E.~Torstensson$^{e,d}$, Kenta Fujisawa$^{g}$ and Alberto Sanna$^a$\thanks{This work was partially funded by the ERC Advanced Investigator Grant GLOSTAR (247078).}\\
       \llap{$^a$} Max-Planck-Institut f\"ur Radioastronomie (MPIfR), Auf dem H\"ugel 69, 53121 Bonn, Germany\\
       \llap{$^b$}Istituto di Fisica dello Spazio Interplanetario (IFSI-INAF), Via del Fosso del Cavaliere 100, 00133 Roma, Italy\\
       \llap{$^c$} Harvard Smithsonian Center for Astrophysics, 60 Garden Street, Cambridge, MA 02138, USA\\
       \llap{$^d$}Joint Institute for VLBI in Europe, Postbus 2, 7990 AA Dwingeloo, the Netherlands\\
       \llap{$^e$}Sterrewacht Leiden, Leiden University, Postbus 9513, 2300 RA Leiden, the Netherlands\\
       \llap{$^f$}Mizusawa VLBI Observatory, National Astronomical Observatory of Japan, 2-21-1, Osawa, Mitaka, Tokio 181-85588,Japan\\
       \llap{$^g$}Faculty of Science, Yamaguchi University, 1677-1 Yoshida,Yamaguchi 753-8512, Japan\\
       E-mail: \email{kazi.rygl@ifsi-roma.inaf.it}, \email{brunthal@mpifr-bonn.mpg.de}, \email{kmenten@mpifr-bonn.mpg.de}, \email{reid@cfa.harvard.edu}, \email{langevelde@jive.nl}, \email{mareki.honma@nao.ac.jp}, \email{kalle@strw.leidenuniv.nl}, \email{kenta@yamaguchi-u.ac.jp}, \email{asanna@mpifr-bonn.mpg.de}}
        
\abstract{Cygnus X is one of the closest giant molecular cloud complexes and therefore an extensively studied region of ongoing high mass star formation. However, the distance to this region has been a long-standing issue, since sources at galactic longitude of ~80$^\circ$ could be in the Local Arm nearby (1--2 kpc), in the Perseus Arm at ~5 kpc, or even in the outer arm (~10 kpc). We use combined observations of the EVN plus two Japanese stations to measure very accurate parallaxes of methanol masers in five star-forming regions in Cygnus X to understand if they belong to one large star-forming complex or if they are separate entities located at different distances. Here we report our preliminary result for W\,75N based on six epochs of VLBI observations: we find that W\,75N is at a distance of $1.32^{+0.11}_{-0.09}$\,kpc, which is significantly closer than the reported values in the literature (1.5--2\,kpc).}

\FullConference{10th European VLBI Network Symposium and EVN Users Meeting: VLBI and the new generation of radio arrays\\
		September 20-24, 2010\\
		Manchester Uk}

\begin{document}

\section{Introduction}
In the early days of radio astronomy, a conspicuously strong, extended 
source of radio emission was found around Galactic longitude $\sim 80^\circ$
and named the Cygnus X region \cite{Pid52}. It stands out equally remarkably 
in infrared (IR) surveys of the Galaxy (\cite{Ode93}; see e.g., the 
spectacular Spitzer imaging \cite{Kum07}). Its associated Giant Molecular
Cloud complex (e.g., \cite{Sch06}) is harboring many dense, dusty, hot cores 
with embedded protostars \cite{Mot07}.  This and the presence of several OB 
associations and a superbubble driven by the O-stars' stellar winds as well 
as the famous Cygnus Loop supernova remnant give testament to intense and 
widespread star formation happening over (at least) the past few millions
of years (see, e.g., \cite{Cas80}; \cite{Abb81}; \cite{Wal55}). Since 
it is also thought to be relatively near to the Sun, many of the above 
phenomena can be studied here in detail. % (for example, it is target
%of multi wavelength campaigns including several key science projects
%of the Herschel satellite). 
The distance to this remarkable ''mini-starburst'' region has been 
a long-standing issue. It is still not clear whether all clouds are at the 
same distance or whether we see a projection of several clouds at different 
distances (see \cite{Sch06} and \cite{Uya01} for a detailed discussion).
For example, three OB associations have distance estimates between 
1.2 and 1.7 kpc, a difference of more than 30\%. This uncertainty of 500\,pc 
is almost 10 times larger than the extent of the Cygnus X region on the sky 
-- 2 by 3$^\circ$ or 50 by 80\,pc at a distance of 1.5\,kpc. Thus, important 
physical parameters of objects in this region such as luminosities and masses \
are  uncertain by a factor of $\sim$2. 
The situation is further complicated by 
the fact that sources at a galactic longitude of $\sim80^\circ$ could be in 
the Local Arm and nearby ($\sim$1$-$2 kpc), in the Perseus Arm at $\sim$5 
kpc, or even in the outer arm at distances of $\sim$10 kpc (e.g. the 
Cygnus X-3 microquasar). 
Fortunately, several 6.7 GHz methanol masers are 
known in the Cygnus X region \cite{Xu09}. %The catalogue of Pestalozzi et al. (2005) lists 
%six masers in this region. %of which five methanol masing regions were detected in the first epochs of EVN observations (EB039A and EB039B, see Table~\ref{tab:maser}). 
One can use this maser line for astrometry; in \cite{Ryg10} we have measured the parallax and proper motions of five star-forming regions (SFRs) and reached accuracies as good as 22\,$\mu$as. 
%for declinations above 20$^\circ$. 
These were the first parallax measurements performed with the European VLBI Network (EVN) .
%(projects EB032 and GB039)  and its accuracies were up to 50 times better than that with the Hipparcos satellite. 
%Here we report the distance ofW\,75N based on the first six epochs. The distances and proper motions of the other maser sources in Cygnus X will be published in a forthcoming article.

\section{Observations and data reduction}

The project EB039 contains eight epochs of observations with the EVN plus two 
Japanese stations (Yamaguchi and Mizusawa) between March, 2009, and November, 
2010, of which six were already correlated and reduced -- the results published 
here are based on these. Each observation lasted 12 hours and made use of 
\emph{ geodetic-like} observations to calibrate the tropospheric zenith delays 
at each antenna (see \cite{Rei04}, \cite{Bru05}, \cite{Rei09} for a detailed 
discussion).

Before we started the parallax observations, we observed several compact NVSS 
\cite{Con98} sources within 2$^\circ$ from the maser source at 5\,GHz with the EVN in eVLBI mode to obtain their positions with sub-arcsecond accuracy. Finally, we used two extragalactic background sources, J2045+4341 and J2048+4310, within 2$^\circ$ of W\,75N and J2029+4636 from the VLBA calibrator survey \cite{Bea02} with a separation of 4.3$^\circ$. A typical observing run started and ended with $\sim$1 hour of
geodetic-like observations and $\sim$10\,minutes of observations of fringe finders. The
remaining time was spent on maser/background source phase-referencing
observations. %During each run, the average on-source time per maser was between $\sim$0.9 and$\sim$1.2\,hours.  
The masers in Cygnus X and the three background sources were phase referenced to the strongest maser in W\,75N, using a switching time of 1.5 minutes. 
 
The observations were performed with eight intermediate frequency bands (IF) of 8\,MHz width, each in dual circular
polarization and  2 bits sampling, yielding a recording rate of 512\,Mbps. The data were correlated
in two passes at the Joint Institute for VLBI in Europe (JIVE). The maser data 
were correlated using one 8\,MHz IF band with
1024 spectral channels, resulting in a channel separation of 7.81\,kHz
or 0.41\,$\mathrm{km~s}^{-1}$ at 6.7\,GHz. The background
sources were correlated in continuum mode with eight IFs of 8\,MHz width
with a channel separation of 0.25\,MHz.   
The data were reduced using the NRAO Astronomical Image Processing System
(AIPS) and ParselTongue \cite{Ket06} following the description in \cite{Ryg10}.
% and included ionospheric delay correction, amplitude calibration, a "manual phase-calibration" for the removal of of delay and phase differences between IFs, correction for the Earth rotation, and Hanning smoothing.

%The data were reduced following the EVN guidelines, applying parallactic angle and
%ionospheric delay corrections. The JIVE correlator model uses Earth's
%orientation parameters, which are interpolated from the appropriate daily-tabulated values,
%so it is not necessary to correct them after the correlation. 
%The ionospheric delays were based on the JPL GPS--IONEX total vertical electron
%content maps of the atmosphere. Amplitudes were
%calibrated using system-temperature measurements and standard gain curves. 
%A ``manual phase-calibration'' was performed to remove delay and phase
%differences between the IFs. The Earth rotation was corrected for with the task `CVEL'. 
%For each maser, a spectral channel with one bright and compact maser spot was
%used as the phase reference. The data was Hanning-smoothed to minimize Gibbs
%ringing in the spectral line data. To avoid the strong fluctuations caused by the
%bandpass edges, the outer two channels in each IF were discarded
%(following \citealt{reid:2009a}). The positions of the masers
%and background sources were extracted by fitting 2D Gaussian to the maps. 

\section{Results and discussion}

\begin{figure}
\centering
\includegraphics[width=12cm]{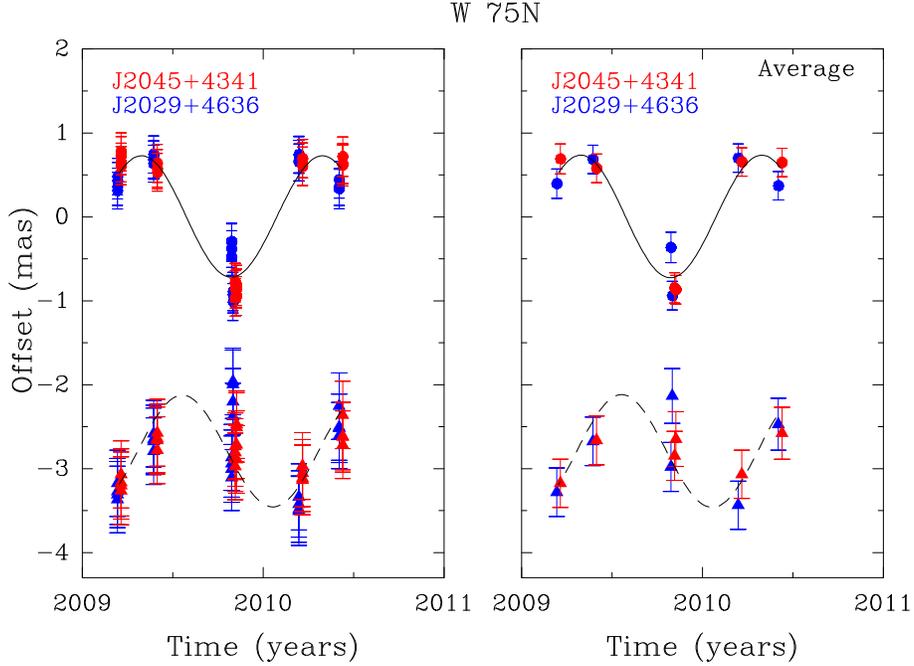}
\caption{Parallax fit to 4 maser spots using 2 background sources (color coded). The left panel shows a fit to all maser spots, and in the right panel the postions of the masers have been averaged per epoch and background source. The solid line represents the fit to the parallax signal in right ascension, while the dashed line represents the fit in declination. Proper motions have already been removed here. }
\label{w75:par} 
\end{figure}

We detected 6.7 GHz methanol masers towards five SFRs: W\,75N, DR\,20, DR\,21A, DR\,21B, and OB\,2.
%For each SFR, we found the emission to arise from a number of separate
%maser spots. Most of the line profiles stretched over
%several channels. Both spatially and/or spectrally different maser spots were
%considered as distinct maser components. Only compact maser spots with well-behaved proper motions were used for the parallax fitting. 
W\,75N has 14 different methanol maser features emitting in a velocity range of 3 -- 10 $\mathrm{km~s^{-1}}$. The parallax  of W\,75N was based on four maser spots which are compact and have well-behaved proper motions, making them suitable for parallax fitting.
%We inspected the behavior of the proper
%motion for each maser spot relative to the reference spot. Maser spots with 
%strong nonlinear proper motions or a large scatter of position about 
%a linear fit were discarded. 
%Only compact maser spots with well-behaved proper motions were used for the parallax fitting. 

%The average internal proper motion of the maser spots ranged between 0.06 and
%$0.23\,\mathrm{mas~yr^{-1}}$. Considering the distance of each maser, these proper
%motions correspond to 0.5--1\,$\mathrm{km~s}^{-1}$, much lower than the
%internal proper motions of water masers, which can reach up to
%20--200\,$\mathrm{km~s}^{-1}$ (\citealt{hachi:2006}). 
%The only exception was ON\,1,
%which separates into two distinct maser groups with a relative proper motion of $0.52\,\mathrm{mas~yr^{-1}}$, or 6.3\,$\mathrm{km~s}^{-1}$. This particularity is
%discussed in Sect.~\ref{sect:on1}.

%\subsection{Fitting of parallax and proper motions}
The parallaxes and proper motions were determined from the change in the
positions of the maser spot(s) relative to the background source(s). The data
were fitted with a parallax and a linear proper motion. 
%Since the formal
%position errors are only based on the signal-to-noise ratios determined from the images,
%they do not include possible systematic errors from residual zenith delay
%errors or source structure changes. This leads to a high reduced $\chi^2$
%value for the fits, so we added error floors in quadrature to 
%the positions until reduced $\chi^2$ values close to unity were reached
%for each coordinate. 
We made combined fits with respect to each background source, assuming one 
parallax but different proper motions for each maser spot. Because the position
measurements of different maser spots are not independent, we multiplied the 
error of the combined fit by $\sqrt N$, where $N$ is the number of maser spots.
However, this will overestimate the error, if significant random errors are
present (e.g., owing to maser blending and structural changes over time),
since the latter are not correlated among different maser spots. Following the 
approach of \cite{Bar08} we calculated the average positions with respect to 
each background source after removing their position offsets and proper 
motions. Then, we performed a parallax fit on these averaged data sets relative
to each individual background source, and on both the background sources 
combined. This approach has the advantage that we can reduce the random errors,
while leaving the systematic errors unaffected.

Figure \ref{w75:par} shows the 6-epoch results for four maser spots in W\,75N 
with respect to two background sources, J2045+4341 and J2029+4636. After 
averaging the maser spots in each epoch, the combined fit to both background 
sources yields a parallax of $0.756\pm0.057$\,mas, corresponding to a 
distance of $1.32^{+0.11}_{-0.09}$\,kpc. This is significantly 
closer than the distance range in the literature of 1.7 to 2.1 kpc 
%(\cite{Red67},\cite{Sch06},\cite{Sur09}) 
which is often based on distance estimates to the Cyg OB associations.

For understanding whether Cygnus X consists of several unconnected SFRs the distances toward all the different SFRs must be measured very precisely. With addition of the two epochs observed this November, and one more year of observations in 2011, we intend to reach accuracies of better than 30\,$\mu$as -- which translates to error bars of 50 pc at a distance of 1.3\,kpc. In addition to the parallaxes we will publish the three-year proper motions of the methanol masers.


\begin{thebibliography}{99}
%\bibitem{...} 
\bibitem{Abb81}Abbott, D.~C., Bieging, J.~H., \& Churchwell, E.\ 1981, \apj, 250, 645 
\bibitem{Bar08}Bartkiewicz, A., Brunthaler, A., Szymczak, M., et al.\ 2008, \aap, 490, 787 
\bibitem{Bea02} Beasley, A.~J., Gordon, D., Peck, A.~B., et al.\ 2002, \apjs, 141, 13
\bibitem{Bru05}Brunthaler, A., Reid, M.~J., \& Falcke, H.\ 2005, Future Directions in High Resolution Astronomy, 340, 455 
\bibitem{Cas80}Cash, W., Charles, P., Bowyer, S., et al.\ 1980, \apjl, 238, L71 
\bibitem{Con98}Condon, J.~J., Cotton, W.~D., Greisen, E.~W., et al.\ 1998, \aj, 115, 169
%\bibitem{Hac09}Hachisuka, K., Brunthaler, A., Menten, K.~M., Reid, M.~J., Hagiwara, Y., \& Mochizuki, N.\ 2009, \apj, 696, 1981 
%\bibitem{Hum78} Humphreys, R.~M.\ 1978, \apjs, 38, 309 
\bibitem{Ket06}Kettenis, M., van Langevelde, H.~J., Reynolds, C., et al. 2006, Astronomical Data Analysis Software and Systems XV, 351, 497 
\bibitem{Kum07}Kumar, M.~S.~N., Davis, C.~J., Grave, J.~M.~C., et al.\ 2007, \mnras, 374, 54
\bibitem{Men91} Menten, K.~M.\ 1991, \apjl, 380, L75
\bibitem{Mot07}Motte, F., Bontemps, S., Schilke, P., et al. 2007, \aap, 476, 1243
\bibitem{Ode93}Odenwald, S.~F., \& Schwartz, P.~R.\ 1993, \apj, 405, 706 
\bibitem{Pes05}Pestalozzi, M.~R., Minier, V., \& Booth, R.~S.\ 2005, \aap, 432, 737 
%\bibitem{Phi05} Phillips, C., \& van Langevelde, H.~J.\ 2005, Future Directions in High Resolution Astronomy, 340, 342 
\bibitem{Pid52}Piddington, J.~H., \& Minnett, H.~C.\ 1952, Australian Journal of Scientific Research A Physical Sciences, 5, 17 
%\bibitem{Red67} Reddish, V.~C.\ 1967, \mnras, 135, 251
\bibitem{Rei04} Reid, M.~J., \& Brunthaler, A.\ 2004, \apj, 616, 872 
\bibitem{Rei09}Reid et al. \ 2009, \apj, 693, 397
\bibitem{Ryg10}Rygl, K.~L.~J., Brunthaler, A., Reid, M.~J., et al.\ 2010, \aap, 511, A2
\bibitem{Sch06}Schneider, N., Bontemps, S., Simon, et al.\ 2006, \aap, 458, 855
%\bibitem{Sur09}Surcis, G., Vlemmings, W.~H.~T., Dodson, R., \& van Langevelde, H.~J.\ 2009, \aap, 506, 757 
\bibitem{Uya01} Uyan{\i}ker, B., F{\"u}rst, E., Reich, W., et al.\ 2001, \aap, 371, 675 
\bibitem{Wal55}Walsh, D., \& Brown, R.~H.\ 1955, \nat, 175, 808 
\bibitem{Xu09}Xu, Y., Voronkov, M.~A., Pandian, J.~D., et al.\ 2009, \aap, 507, 1117

\end{thebibliography}
\end{document}